\documentclass[final,aps,prd,twocolumn,showpacs,showkeys, 10pt, longbibliography, nolinenumbers, floatfix]{revtex4-2}

\usepackage[T1]{fontenc}
\usepackage{amsmath}
\usepackage{amssymb}
\usepackage{amsfonts}
\usepackage{graphicx}
\usepackage{float}
\usepackage{dcolumn}
\usepackage{multirow}
\usepackage{natbib}
\usepackage[colorlinks = true,linkcolor = blue,urlcolor  = blue,citecolor = blue,anchorcolor = blue]{hyperref}
\usepackage{enumerate}
\usepackage{times}
\usepackage{microtype}

\begin{document}

\title{Configurational entropy as a probe of the stability condition of compact objects}

\author{P.S. Koliogiannis}
\email{pkoliogi@physics.auth.gr}
\author{G.A. Tsalis}
\author{C.P. Panos}
\author{Ch.C. Moustakidis}
\email{moustaki@auth.gr}

\affiliation{Department of Theoretical Physics, Aristotle University of Thessaloniki, 54124 Thessaloniki, Greece}

\begin{abstract}

We systematically extend the statement  that the configurational entropy provides an alternative approach to studying gravitational stability of compact objects, carried out in the previous work of M. Gleiser and N. Jiang, Phys. Rev. D {\bf 92}, 044046 (2015). Inspired  by that paper, we try to answer the crucial question: Is there any one-to-one correspondence between the minimum of the configurational entropy  and the stability point for each realistic equation of state? In view of the above question,  we focus on neutron stars, quark stars, as well as on  the third family of compact stars (hybrid stars), where a possible phase transition may lead to the existence of twin stars (stars with equal mass but different radius). In each case, we use a large set of equations of state investigating the possibility to find correlations between the stability region obtained from the traditional perturbation methods to the one obtained by the minimum of configurational entropy. We found that the suggested prediction of the stability by the minimization of the configurational entropy, concerning neutron stars and quark stars, does not have, at least quantitatively, universal validity. However, in several cases it qualitatively predicts the existence of the stability point. In conclusion, the configurational entropy, can be considered  as  an additional tool which enters into the quiver for studying the stability of compact objects such as for example neutron and quark stars, but taking into account that the accuracy of the predictions depends on the specific character of each  equation of state.


\end{abstract}

\maketitle
\section{Introduction}

The concept of the configurational entropy (CE) has been introduced by Gleiser and Stamatopoulos~\cite{Gleiser-2012} in order to study the possible relation of the dynamical and information content of physics models to localized energy configurations (see also  relevant applications in 
Refs.~\cite{Gleiser-2013,Gleiser-2015a,Gleiser-2015b,Braga-2019,Braga-2020,Yunes-2019,Rocha-2021,Karapetyan-2018,Correa-2016,Gleiser-2018}). Gleiser and Jiang~\cite{Gleiser-2015a} trying to investigate the possibility to obtain information about the stability of compact objects from its information-entropic measure, found that the configurational entropy offers an alternative way to compute the critical mass for a variety of stellar objects. In particular, they applied their approach mainly in white dwarfs and boson stars, but also in neutron stars by applying the case of the Fermi gas equation of state (EOS). They found that the traditional perturbation methods correlate well with critical points of the configurational entropy with an accuracy of a few percent or better~\cite{Gleiser-2015a}.

The stability of relativistic stars is a longstanding problem. There are various methods to examine the stability of a relativistic star (for a detailed presentation see Ref.~\cite{Bardeen-1966}). However, in the present work, in addition to the information-entropic measure, we employ the method  based on the dependence of the gravitational mass $M$ and the corresponding radius $R$ on the central energy density ${\cal E}_c$. The stability condition demands that the mass increases with increasing central energy density $dM/d{\cal E}_c>0$. The extrema in the mass indicates a change in the stability of the compact star configuration~\cite{Haensel-2007,Bielich-2020}.

It is worth noticing that there is not any theoretical argument (or proof) to relate the stability point to the minimum of the configurational entropy. However, it is intuitive to expect that since the maximum mass corresponds to the most compact configuration (maximum mass and minimum radius of a stable configuration), the corresponding CE will exhibit an extreme value (in this case a total minimum). The crucial question that needs to be answered is: Is there any  one-to-one correspondence between the minimum of the CE and the stability point for each realistic EOS? And even more: Is this rule universal or it depends on the specific character of each EOS? If this is true, then an additional tool enters into the  quiver for studying the stability of compact objects. 

The motivation of the present work is to provide an extended examination of the results  of Ref.~\cite{Gleiser-2015a}. To this end, we apply two analytical solutions of the Tolman-Oppenheimer-Volkoff (TOV) equations~\cite{Haensel-2007}  and also  a large set of realistic EOSs in order to study the dependence of the bulk neutron star properties on the configurational entropy, while inversely relating our focus, mainly, on the stability conditions. In order to enrich our systematic study, we also use a set of EOSs corresponding to interacting quark matter (suitable to describe the properties of a quark star). Finally, we apply a set of EOSs, with a more complex character, constructed to reproduce properties of a third family of compact objects, known as twin stars. Thus, we consider that we have significantly ensured the examination of the findings of Ref.~\cite{Gleiser-2015a}, covering a very large range of cases which corresponds to compact objects.

The paper is organized as follows. In Sec.~\ref{sec:conf-eos} we present the basic formalism of the configurational entropy and we provide the  realistic EOSs used in the present work. In Sec.~\ref{sec:results} the results of the present study are presented and discussed. Finally, concluding remarks are given in Sec.~\ref{sec:remarks}.

\section{Configurational entropy  and equations of state }
\label{sec:conf-eos}
The key quantity to calculate the configurational entropy in momentum space is the Fourier transform $F({\bf k})$ of the density $\rho(r)={\cal E}(r)/c^2$ (where $\mathcal{E}$ is the energy density), originating from the solution of the TOV equations, that is~\cite{Gleiser-2015a}
\begin{eqnarray}
F({\bf k})&=&\int\int\int \rho(r)e^{-i {\bf k}\cdot {\bf r}} d^3{\bf r}.
\label{Fk-1}
\end{eqnarray}
Moreover, the modal fraction $f({\bf k})$ is defined as
\begin{equation}
f({\bf k})=\frac{|F({\bf k})|^2}{\int|F({\bf k})|^2d^3{\bf k}},
\label{Fk-2}
\end{equation}
and the normalized one as $\tilde{f}({\bf k})=f({\bf k})/f({\bf k})_{\rm max}$, where $f({\bf k})_{\rm max}$ is the maximum fraction, which is given in many cases by the zero mode $k=0$, or by the system's longest physics mode, $|k_{min}|=\pi/R$. The above procedure guarantees that $\tilde{f}({\bf k}) \leq 1$ for all values of ${\bf k}$. We note here that the system's longest physics mode, $|k_{min}|=\pi/R$, is the one that has been used throughout the present work. Finally, the configurational entropy $S_C$, as a functional of $\tilde{f}({\bf k})$, is given by
\begin{equation}
S_C[\tilde{f}]=-\int \tilde{f}({\bf k}) \ln [\tilde{f}({\bf k})] d^3{\bf k}.
\label{S-1}
\end{equation}

In detail, we use a set of hadronic EOSs, which have been extensively employed in the literature for applications in neutron star properties (see Ref.~\cite{Koliogiannis-2020} and references therein). These EOSs have been collected in order to satisfy the prediction of the maximum observed neutron star mass, that is $M \geq 2 M_{\odot}$. In this case, for a specific EOS and for each M-R configuration, we implement the corresponding density distribution $\rho(r)$. Afterwards, the calculation of $F(k)$, $f(k)$, and $\tilde{f}(k)$ is performed according to the presented recipe. It has to be mentioned that it is well established that the stability point corresponds to the configuration of the maximum mass~\cite{Haensel-2007}.

Furthermore, we also employ a set of EOSs concerning interacting quark matter which has been predicted and applied in Ref.~\cite{Zhang-2021}. In this case, the pressure is related to the energy density via the simplified expression~\cite{Zhang-2021}
\begin{eqnarray}
 \frac{P}{4 B_{\rm eff} }&=&\frac{1}{3}\left( \frac{\cal E}{4 B_{\rm eff} }-1  \right)  \\
 &+&
 \frac{4}{9\pi^2}\bar{\lambda}\left( -1+\sqrt{1+\frac{3\pi^2}{\bar{\lambda}}\left( \frac{\cal E}{4 B_{\rm eff} }-\frac{1}{4} \right)  }  \right), \nonumber
\end{eqnarray}
where $\mathcal{E}$ denotes the energy density and  $P$ is the pressure.
Specifically, in the present work we employed the value of $B_{\rm eff}=150~{\rm MeV~fm^{-3}}$, while for the dimensionless parameter $\bar{\lambda}$, we applied the values of $(1,2,5,16)$~\cite{Zhang-2021}.

Finally, we employed a more complex EOS, constructed to reproduce properties of a third family of compact objects, known as twin stars~\cite{Bielich-2020,Alford-2013}. To be more specific, we employed the Maxwell construction, suitable to describe phase transitions in the interior of a compact object and formulated as follows~\cite{Alford-2013}
\begin{equation}
  \mathcal{E}(P) = \begin{cases} 
      \mathcal{E}(P), & P\leq P_{\rm tr} \\
      \mathcal{E}(P_{{\rm tr}}) + \Delta \mathcal{E} + c_s^{-2}(P-P_{{\rm tr}}), & P \geq P_{{\rm tr}}.
   \end{cases}
   \label{MC-1}
\end{equation}
In the above formula, $c_{\rm s}=\sqrt{{\partial P}/{\partial {\cal E}}}$ is the speed of sound (in units of the speed of light), and $\Delta \mathcal{E}$ is the magnitude of the energy density jump at the transition point. The subscript $``{\rm tr}"$ denotes the corresponding quantity at this point. In the region $P\leq P_{\rm tr}$,  we utilized the GRDF-DD2 EOS~\cite{Typel-2018,Tsaloukidis-2023} while in the region $P\geq P_{\rm tr}$, the value of the speed of sound is fixed at $c_{\rm s}=1$. 

\begin{figure}
\includegraphics[width=0.875\columnwidth]{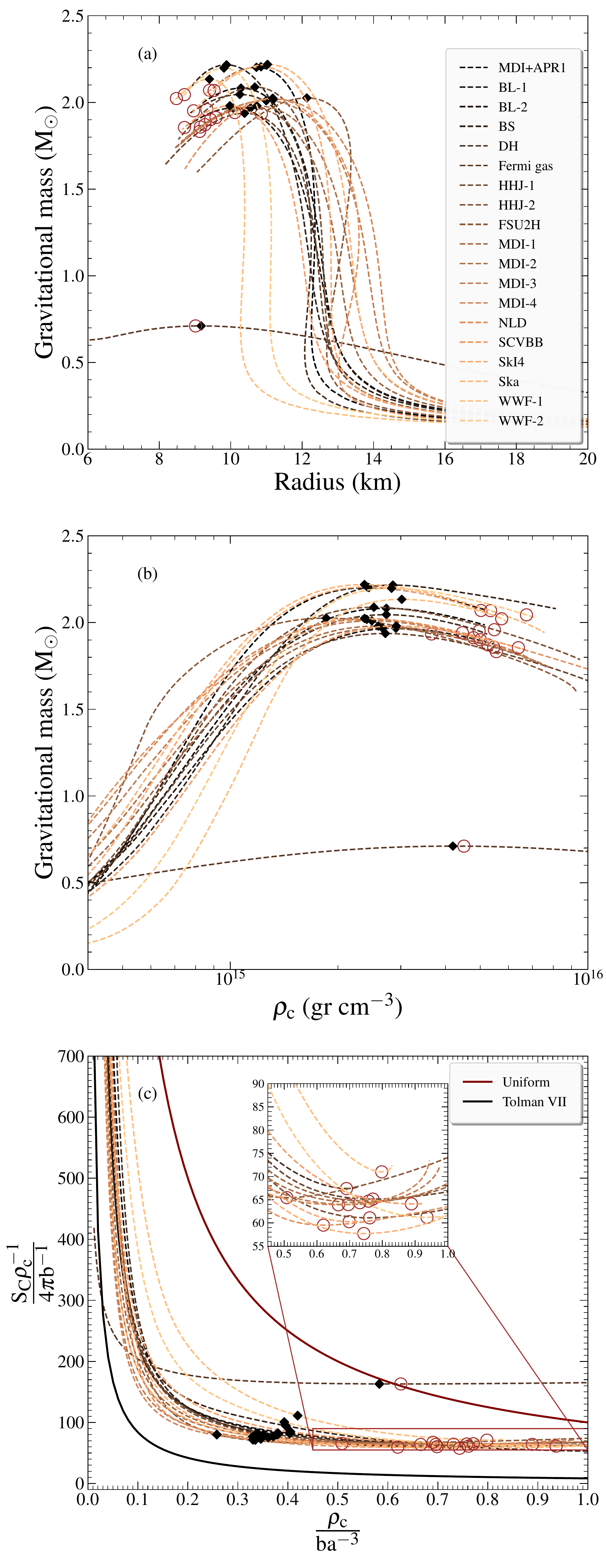}
\caption{(a) Gravitational mass as a function of the radius for a set of hadronic EOSs. (b) The corresponding dependence of the gravitational mass as a function of the central density. (c) The corresponding configurational entropy as a function of the central density and two analytical solutions of TOV equations (${\rm a}$ and ${\rm b}$ are constants~\cite{Gleiser-2015a}). The inset figure indicates the location of the minimization of the CE. The black diamonds indicate the stability points due to the TM while the open circles correspond to the minimum of the CE. The hadronic EOSs are presented with the dashed lines, and the analytical solutions with the solid ones.
}
\label{fig:hadronic_eos}
\end{figure}

\begin{figure}
\includegraphics[width=0.875\columnwidth]{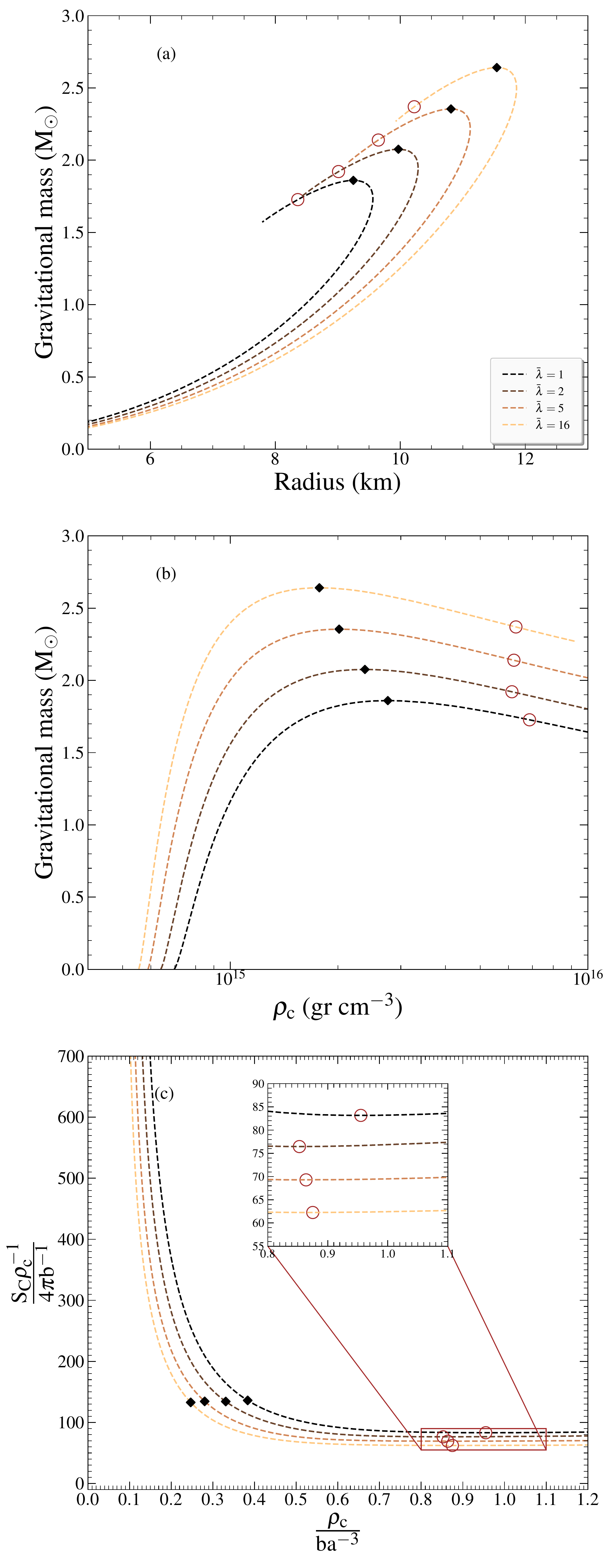}
\caption{(a) Gravitational mass as a function of the radius for a set of quark star EOSs. (b) The corresponding dependence of the gravitational mass as a function of the central density. (c) The corresponding configurational entropy as a function of the central density. The inset figure indicates the location of the minimization of the CE. The black diamonds indicate the stability points due to the TM while the open circles correspond to the minimum of the CE.}
\label{fig:quark_eos}
\end{figure}

\begin{figure}
\includegraphics[width=0.875\columnwidth]{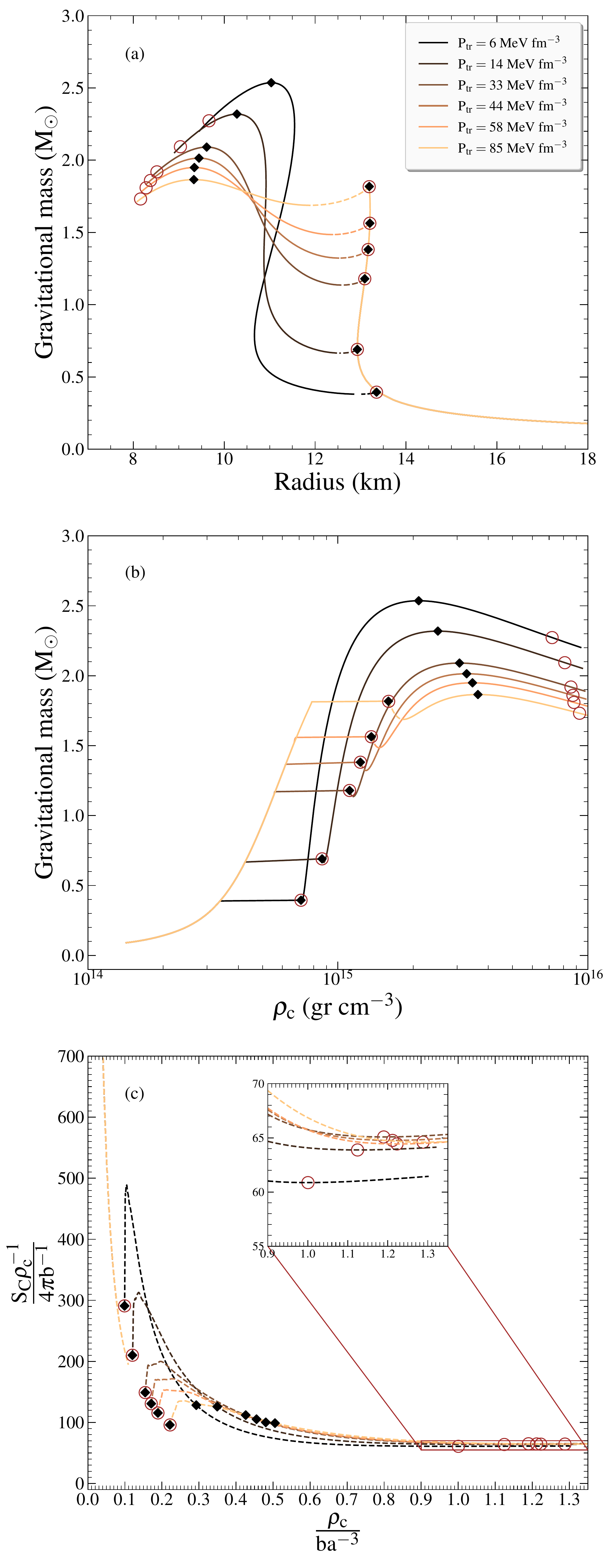}
\caption{(a) Gravitational mass as a function of the radius for a set of twin-star EOSs. (b) The corresponding dependence of the gravitational mass as a function of the central density. (c) The corresponding configurational entropy as a function of the central density. The inset figure indicates the location of the minimization of the CE. The black diamonds indicate the stability points due to the TM while the open circles correspond to the minimum of the CE. In (a) and (b) the twin star EOSs are presented with the solid lines, where the dashed lines represent the intermediate unstable region, and in (c) the corresponding EOSs are presented with the dashed lines.}
\label{fig:twin_star_eos}
\end{figure}

\section{Results and Discussion}
\label{sec:results}
In Fig.~\ref{fig:hadronic_eos}(a) we display the dependence of the gravitational mass on the radius for the hadronic EOSs, including the Uniform and Tolman VII analytical solutions for $R=12$ km~\cite{Kramer-1980}. The diamonds indicate the stability points due to the traditional method (TM, which corresponds to the maximum mass configuration), while the open circles correspond to the minimum of the CE. It is obvious that in all cases, the location of the stability point does not coincide with the one corresponding to the minimum of the CE. Moreover, in some cases, no configuration appears to minimize CE, even at high values of central density, far from the stability point. In order to further clarify this point, in Fig.~\ref{fig:hadronic_eos}(b) we indicate the dependence of the mass on the central density. The non-coincidence of the corresponding stability points is evident in this case as well. Finally, in Fig.~\ref{fig:hadronic_eos}(c) we present the CE as a function of the central density. In all cases, we found that the configurational entropy is a decreasing function of the central density, where the appearance of the minimization is at high-values of density (if it exists). Therefore, the minimization point of CE and the maximum mass configuration of TM do not coincide, since the first one is located at high-values of density, far from those corresponding to the stability point. It has to be noted that the high-values of density that correspond to the minimization of CE, denote neutron stars in the instability region. However, there is only one case where the two mentioned points nearly coincide. This limiting case is the free Fermi gas which is shown in Fig.~\ref{fig:hadronic_eos}.

\begin{figure}
    \includegraphics[width=0.875\columnwidth]{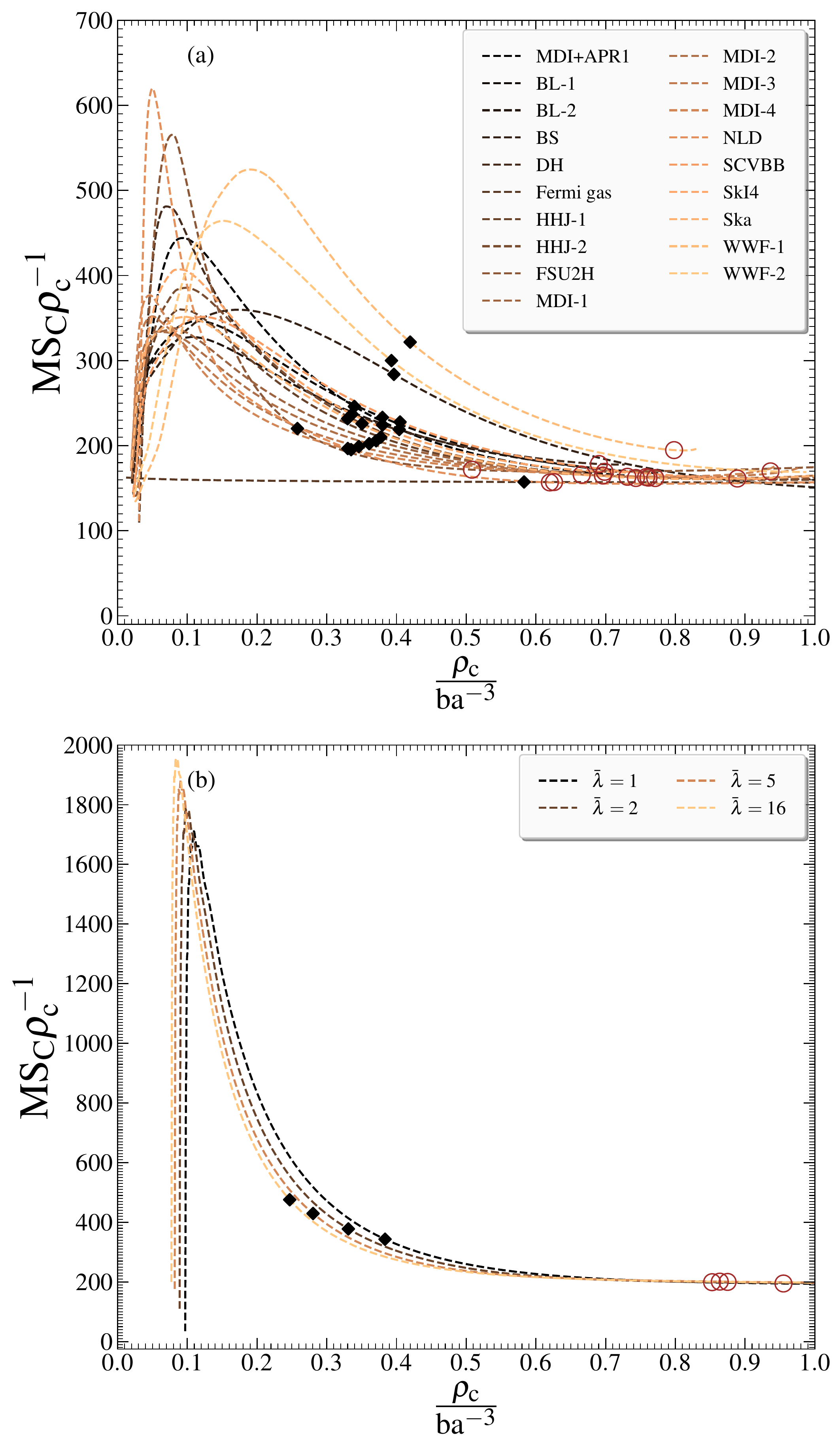}
    \caption{(a) The $MS_{\rm C}\rho_{c}^{-1}$ as a function of the central density for (a) hadronic neutron stars and (b) quark stars. The black diamonds indicate the stability points due to the TM while the open circles correspond to the minimum of the CE.}
    \label{fig:hadrons_quarks_eos}
\end{figure}

Quark stars have different configurations compared to neutron stars, since at the surface while the pressure is vanishing, that is not the case for the density. This behavior is well reflected on the dependence of the gravitational mass on the radius. In this case too, for a set of 4 EOSs, the predictions confirm the discrepancy between TM and CE stability points, as shown in Fig.~\ref{fig:quark_eos}.
Finally, in the scenario of twin stars (hybrid stars), two stability points exist for the 6 EOSs under consideration. From Fig.~\ref{fig:twin_star_eos} the first stability point coincides with the first minimum of the CE (less than 2$\%$ error from the stability point). However, the latter could be justified, since at this point a discontinuity on the EOS manifests as
a large gap in the central energy density. Consequently, this minimum can be identified as an artificial one. Regarding the  second minimum of CE, in this case too, it appears very far from the second stability point (more than two to three times the central density).

Moreover, in order to further clarify the connection between the minimization of the CE and the stability point (which corresponds to the maximum mass configuration), we display in Fig.~\ref{fig:hadrons_quarks_eos} the dependence of $MS_{\rm C}\rho_{c}^{-1}$ as a function of the central density $\rho_{c}/ba^{-3}$ for (a) hadronic neutron stars, and (b) quark stars. It is worth noticing that while in the case of the Newtonian polytropes the quantity $MS_{\rm C}\rho_{c}^{-1}$ is nearly constant along the $\rho_{c}/ba^{-3}$ values (see also Ref.~\cite{Gleiser-2015a}), this rule is not satisfied, at least at low densities, both in neutron stars and quark stars (the only exception is the case of Fermi gas in accordance to the prediction of Ref.~\cite{Gleiser-2015a}). We consider that in the majority of the cases, the more complex dependence of $MS_{\rm C}\rho_{c}^{-1}$ on $\rho_{c}/ba^{-3}$ is mainly due to the special characteristics of each EOS. To be more specific, for low values of the density, there is a dramatic divergence from linearity, thus reflecting the specific feature of each EOS in this region. On the contrary, at high densities the linearity seems to be restored. We guess that this is the reason why the two different predictions of the stability points agree qualitatively but not quantitatively.

It is worth noticing that the existence of a stable configuration is a property of gravity and does not depend on the EOS. However, the specific location of the stability point depends on the EOS~\cite{Glendenning-2000b}. In view of the above comments, ones expects, according to the statement of Ref.~\cite{Gleiser-2015a}, the minimization of the CE to be a consequence of gravity (in the framework of  general relativity). If this is true, then we expect the existence of the relevant minimum to be independent of the applied EOS. Nonetheless, we found that this is not a general rule for the case of neutron stars and quark stars. To be more specific, while there are cases where the CE is minimized, there are also cases where the CE is monotonically decreasing, with no appearance of a minimum. In addition, even in the first case scenario, the location of the minimum does not coincide with the stability point. In spite of the latter, there is only one exception, which is the free Fermi gas.

\section{Concluding Remarks}
\label{sec:remarks}
Summarizing, our results lead to the conclusion that the CE can  be used partially as a measure of the stability of compact objects including neutron stars and quark stars.  We could speculate that CE can only qualitatively predict, in some way, some configurations, which can be correlated to the stability points. Concluding,  although the approach of minimization of the CE may be fascinating as an alternative way to study the stability point of neutron stars and quark stars, it is mainly qualitative and not quantitative in the sense  that the results strongly depend  on the applied EOS.

Finally, it is necessary to state that, in the present work, we have not exhausted the investigation of the correlation of CE to the stability conditions of compact objects. A future perspective is to consider the above correlation in the case of rapidly rotating or anisotropic stars (where the pressure is anisotropic). It will be very interesting to investigate and even more to  reveal possible correlation between stability and CE in the mentioned cases. This can only be done when the corresponding research is carried out in every detail and covers all possible cases.

\section*{Acknowledgments}
The authors would like to thank Dr. Nan Jiang for the correspondence and useful comments. The implementation of the research was co-financed by Greece and the European Union (European Social Fund-ESF) through the Operational Programme \guillemotleft Human Resources Development, Education and Lifelong Learning\guillemotright ~in the context of the Act ``Enhancing Human Resources Research Potential by undertaking a Doctoral Research” Sub-action 2: IKY Scholarship Programme for PhD candidates in the Greek Universities. All numerical calculations were performed on a workstation equipped with 2 Intel Xeon Gold 6140 Processors (72 cpu cores in total) provided by the MSc program “Computational Physics” of the Physics Department, Aristotle University of Thessaloniki.


\end{document}